\documentclass[12pt]{iopart}

\usepackage{graphicx} 
\begin{document}

\title{Strongly coupled matter near phase transition}
\author{Bao-Chun Li$^{1,2}$, Mei Huang$^{2,3}$} 
\address{$^{1}$ Institute of Theoretical Physics, Shanxi
University,Taiyuan Shanxi, China\\
$^{2}$ Institute of High
Energy Physics, Chinese Academy of Sciences, Beijing, China\\
$^{3}$ Theoretical Physics Center for Science Facilities, Chinese Academy of Sciences, Beijing, China }

\begin{abstract}
In the Hartree approximation of Cornwall-Jackiw-Tomboulis (CJT) formalism of the real scalar 
field theory, we show that for the strongly coupled scalar system near phase transition, the shear 
viscosity over entropy density is small, however,  the bulk viscosity over entropy density is large. 
The large bulk viscosity is related to the highly nonconformal equation of state. It is found that the 
square of the sound velocity near phase transition is much smaller than the conformal value $1/3$, 
and the trace anomaly at phase transition  deviates far away from $0$. These results agree well 
with the lattice results of the complex QCD system near phase transition. 
\end{abstract}

\maketitle

It has been believed that quark matter created at RHIC behaves like a nearly ``perfect" fluid.
One crucial quantity to identify this property is the small ratio of shear viscosity over 
entropy density $\eta/s$. In order to fit the elliptic flow
at RHIC, the hydrodynamic simulation shows that a very small shear viscosity is required 
\cite{Hydro}. Lattice QCD calculation confirmed that $\eta/s$ for the purely gluonic plasma 
is rather small and in the range of $0.1-0.2$ \cite{LAT-etas}. 

However, recent lattice QCD results showed that another important transport coefficient, 
the bulk viscosity over entropy density ratio $\zeta/s$ rises dramatically up to the order 
of $1.0$ near the critical temperature $T_c$ \cite{LAT-xis-KT,LAT-xis-Meyer}. The sharp peak 
of bulk viscosity at $T_c$ has also been observed in the linear sigma model \cite{bulk-Paech-Pratt}, 
and the increasing tendency of $\zeta/s$ below $T_c$ has been shown in a massless pion gas
\cite{bulk-Chen}. The large bulk viscosity near phase transition is related to the 
nonconformal equation of state \cite{LAT-EOS}. The correlation between large
bulk viscosity and the nonconformal euqation of state has  been shown in Ref. \cite{bulk-Nicola}.

The bulk viscosity is related to the correlation function of the trace of the 
energy-momentum tensor $\theta^\mu_\mu$:
\begin{equation}
\label{kubo}
\zeta = \frac{1}{9}\lim_{\omega\to 0}\frac{1}{\omega}\int_0^\infty dt \int d^3r\,e^{i\omega t}\,\langle [\theta^\mu_\mu(x),\theta^\mu_\mu(0)]\rangle \,.
\end{equation}
According to the result derived from low energy theorem, in the low frequency region, the bulk viscosity takes the form of ~\cite{LAT-xis-KT}
\begin{eqnarray}\label{ze}
\,\zeta = \frac{1}{9\,\omega_0}\left\{ T^5\frac{\partial}{\partial T}\frac{(\epsilon_T-3p_T)}{T^4}+16|\epsilon_v|\right\}\,
  =  \frac{1}{9\,\omega_0} \left\{- 16 \epsilon+9 T s + T C_v \right\}\,.
\end{eqnarray}
Here $\epsilon$ is the total energy density of the system, $\epsilon_v$
the negative vacuum energy density,  $s$ is the entropy density and $C_v$ is the specific heat. 
We have introduced the normalized pressure density $p_T$ and energy density $\epsilon_T$.
The parameter $\omega_0 = \omega_0(T)$ 
is a scale at which the perturbation theory becomes valid. 

There are several quantities to characterize the confomality of the system. One is the square 
of the speed of sound $c_s^2$, which is related to $p_T/\epsilon_T$ and has the form of
\begin{equation}
c_s^2=\frac{{\rm d}p}{{\rm d}\epsilon}=\frac{s}{T {\rm d}s/{\rm d}T}=\frac{s}{C_v}.
\end{equation}
At the critical temperature, the entropy density as well as 
energy density change most fastly with temperature, thus one expect that $c_s^2$
should have a minimum at $T_c$. 
Another quantity to describe the conformality of the system is the trace anomaly of the 
energy-momentum tensor ${\cal T}^{\mu\nu}$, i.e,
\begin{equation}
\Delta=\frac{{\cal T}^{\mu\mu}}{T^4}\equiv \frac{\epsilon_T-3 p_T}{T^4}
=T\frac{\partial}{\partial T}(p_T/T^4).
\end{equation}
We define $\Delta/d$ as the "interaction measure", with $d$ the degeneracy factor. 

The conformal limit has attracted much 
attention in recent years, since people are trying to understand strongly 
interacting quark-gluon plasma by using AdS/CFT techniques. 
In conformal field theories including free field theory, 
$p_T/\epsilon_T=c_s^2=1/3$, $\Delta=0$,
and the bulk viscosity $\zeta$ is always zero. 
Lattice results show that at asymptotically high temperature,
the hot quark-gluon system is close to a conformal and free ideal gas. 

However, lattice results show that near deconfinement phase transition, 
the hot quark-gluon system deviates far away from conformality \cite{LAT-EOS}. 
Both $p_T/\epsilon_T$ and $c_s^2$ show a minimum  around $0.07$, which  
is much smaller than $1/3$. For the $SU(3)$ pure gluon system,
the peak value of the trace anomaly $\Delta_{LAT}^{G}$ reads $3\sim 4$ at
$T_{max}$ and the corresponding "interaction measure" is 
$\Delta_{LAT}^{G}/d_G=0.2\sim 0.25$, with the gluon degeneracy factor $d_G=16$.
(Note that here $T_{max}\simeq 1.1 T_c$ is the temperature corresponding to the sharp peak 
of $\Delta$.) For the two-flavor case, the lattice result of the 
peak value of the trace anomaly $\Delta_{LAT}^{Nf=2}$ reads $8\sim 11$, 
the corresponding interaction measure at $T_{max}$ is given as 
$\Delta_{LAT}^{Nf=2}/(d_G+d_Q)=0.28 \sim 0.4$, with quark degeneracy factor $d_Q=12$. 

Due to the complexity of QCD in the regime of strong coupling, results on hot 
quark matter from lattice calculation and hydrodynamic simulation 
are still lack of analytic understanding. 
In recent years, the anti-de Sitter/conformal field theory (AdS/CFT) 
correspondence has generated enormous interest in using thermal ${\cal N} = 4$ 
super-Yang-Mills theory (SYM) to understand sQGP. 
The shear viscosity to entropy density ratio $\eta/s$ is as small as $1/4\pi$ in the strongly 
coupled SYM plasma \cite{etaovers-AdS}.
However, a conspicuous shortcoming of this approach is the conformality of SYM:
the square of the speed of sound $c_s^2$ always equals to $1/3$ and the bulk viscosity is 
always zero at all temperatures in this theory. Though $\zeta/s$ at $T_c$ is non-zero
for a class of black hole solutions resembling the equation of state of QCD,  the magnitude is 
less than $0.1$ \cite{Gubser-EOS}, which is too small comparing with lattice QCD results.

It has been found in Ref. \cite{etas-scalar} that in the simplest real scalar model with 
$Z(2)$ symmetry breaking in the vacuum, $\eta /s$ behaves the same way as that in 
systems of water, helium and nitrogen in first-, second-order phase transitions and 
crossover \cite{Csernai:2006zz}. In Ref. \cite{bulk-Li-Huang}, we have investigated the equation 
of state and bulk viscosity in the real scalar model, and compare the result in this simplest
relativistic system with that of the complex QCD system.

The Lagrangian of the real scalar field theory has the form of
\begin{equation}
\mathcal{L}=\frac{1}{2}(\partial _{\mu }\phi )^{2}-\frac{1}{2}a\phi ^{2}-%
\frac{1}{4}b\phi ^{4},
\end{equation}%
with $a$ the mass square term and $b$ the interaction strength. 
This theory is invariant under $\phi \rightarrow -\phi $ and has a $Z_{2}$
symmetry. In the case of $a<0$ and $b>0$, the vacuum at $T=0$ breaks the $Z_{2}$ 
symmetry spontaneously. The $Z(2)$ symmetry will be restored at finite temperature
with a second-order phase transition. 
  
At finite temperature, the naive perturbative expansion in powers of the coupling 
constant breaks down. A convenient resummation method is provided by 
the extension of Cornwall-Jackiw-Tomboulis (CJT) formalism \cite{CJT} to finite 
temperature. The CJT formalism is equivalent to the $\Phi$-functional approach of 
Luttinger and Ward \cite{Luttinger} and Baym \cite{Baym}. In our calculation, we only 
perform the Hartree approximation for the effective potential, i.e, only resum tadpole
diagrams self-consistently and neglect the exchange diagrams. 
The effective potential in the CJT formalism reads \cite{CJT-Dirk} 
\begin{eqnarray}
\Omega[\bar{\phi},S] =\frac{1}{2}\int_{K}\left[ \,\ln
S^{-1}(K)+S_{0}^{-1}(K)\,S(K)-1\,\right] 
+\,\,V_{2}[\bar{\phi},S]+U(\bar{\phi})\,\,,
\end{eqnarray}%
where $U(\bar{\phi})=a/2~\bar{\phi}^{2}+b/4~\bar{\phi}^{4}$ is the 
tree-level potential, and the 2PI potential 
$V_{2}[\bar{\phi},S]=\frac{3}{4} b \left(\int_{K}\,S(K,\bar{\phi})\right)^2$ 
in the Hartree approximation.
$S(S_{0})$ is the full(tree-level) propagator and takes the form of
$S^{-1}(K,\bar{\phi})=-K^{2}+m^{2}(\bar{\phi}),~
S_{0}^{-1}(K,\bar{\phi})=-K^{2}+m_{0}^{2}(\bar{\phi})$ with
the tree-level mass $m_{0}^{2}=a+3b~\bar{\phi}^{2}$.

The gap equations for the condensation $\phi _{0}$ and scalar mass $m$ 
are determined by the self-consistent one- and two-point Green's functions
\begin{equation}
\left. \frac{\delta \Omega}{\delta \bar{\phi}}\right\vert _{\bar{\phi}=\phi
_{0},S=S(\phi _{0})}\equiv 0\;,\;\;\;\left. \frac{\delta \Omega}{\delta \bar{S}}%
\right\vert _{\bar{\phi}=\phi_0 ,S=S(\phi _{0})}\equiv 0.
\end{equation}

The entropy density is determined by taking the derivative of effective potential 
with respect to temperature, i.e, $s=-\partial \Omega(\phi_0)/\partial T$. 
In the symmetry breaking case, the vacuum effective potential or the vacuum energy density
is negative, i.e, $\Omega_{v}=\Omega(\phi_0)|_{T=0}<0$. The normalized energy density $\epsilon_T$
and pressure density $p_T$ can be calculated by $p_T=-\Omega_T$ with 
$\Omega_T=\Omega(\phi_0)-\Omega_{v}$ and $\epsilon_T=-p T +Ts $.

\begin{figure}[htp]
\centering
\includegraphics[width=76mm]{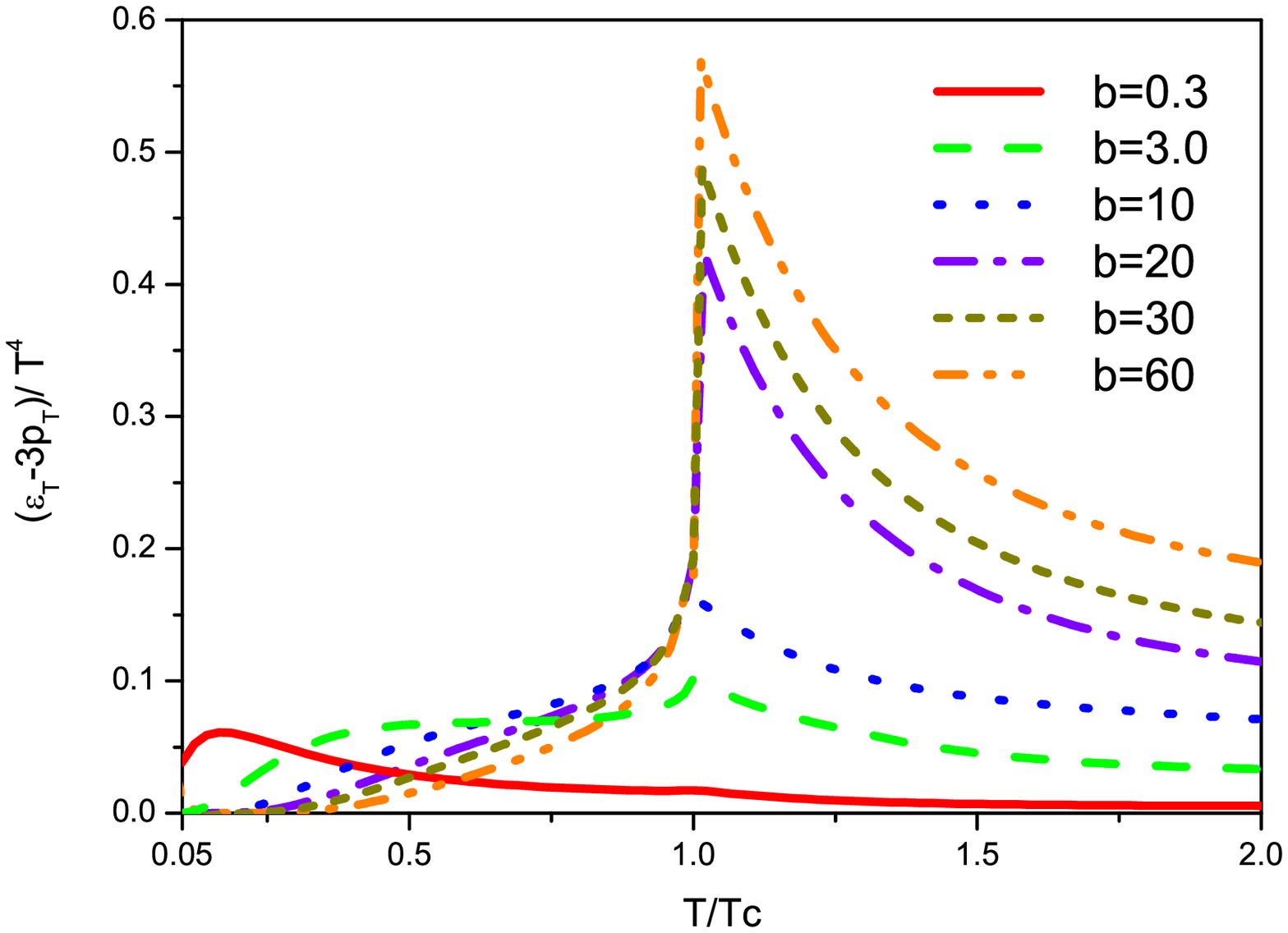}
\hfill
\includegraphics[width=76mm]{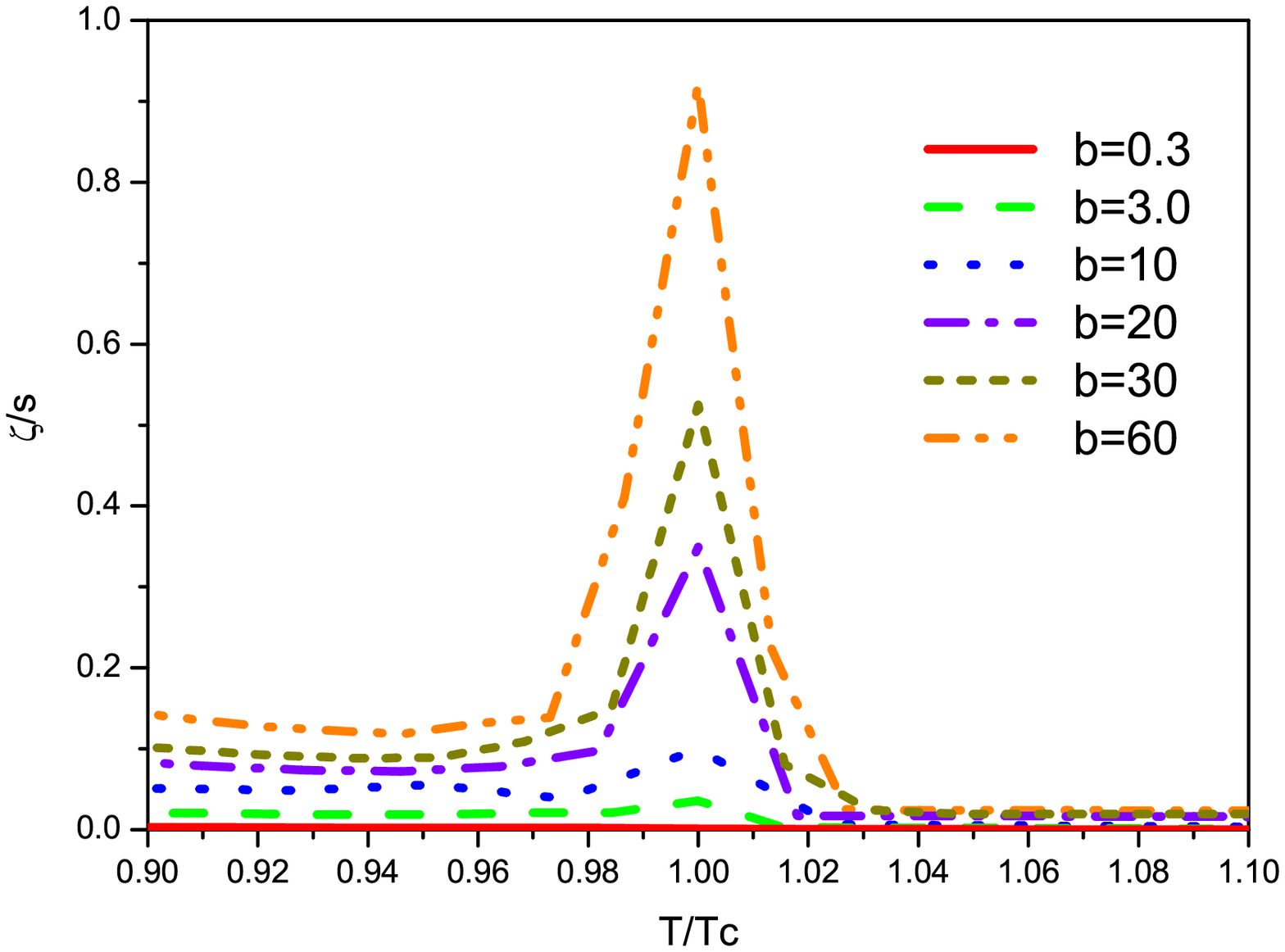} 
\caption{\textit{The interaction measure $(\epsilon_T-3 p_T)/T^4$ (left) and the bulk viscosity over entropy density ratio $\zeta/s$ (right) as a function of temperature $T/T_c$ for different coupling 
strength $b$. }}%
\label{intmeas-bulk-fig}%
\vspace*{-0.15cm}
\end{figure}

As a reference for complex QCD system, we investigate the equation of state and transport 
properties for the real scalar field theory with $Z(2)$ symmetry breaking in the vacuum and 
2nd order phase transition at finite temperature. 
The trace anomaly $\Delta$， the specific heat $C_v$ as well as bulk viscosity to entropy 
density ratio $\zeta/s$ show upward cusp at $T_c$, and their peak values increase with 
the increase of coupling strength. The ratio of pressure density over energy density 
$p_T/\epsilon_T$ and the square of the sound velocity $c_s^2$ show downward cusp at $T_c$,
which is similar to the behavior of $\eta/s$ found in Ref. \cite{etas-scalar}, and
the cusp values decrease with the increase of coupling strength. 
These cusp behaviors at phase transition resemble lattice QCD results. 
In Fig. \ref{intmeas-bulk-fig}, we only show the trace anomaly $\Delta$
and the ratio of bulk viscosity to entropy density ratio $\zeta/s$ as 
functions of $T/T_c$ for different coupling strength $b$. 

In the weak coupling case when $b=0.3$, the cusp values of $p_T/\epsilon_T$ and $c_s^2$
at ${T_c}$ are close to the conformal value $1/3$, both the trace anomaly $\Delta$ 
and the bulk viscosity to entropy density ratio $\zeta/s$ at $T_c$ are close to conformal 
value $0$. However, the shear viscosity over entropy density ratio $\eta/s$ is around $2000$,
which is huge comparing with the AdS/CFT limit $1/4\pi$. Here we have used the method in
Ref.\cite{etas-scalar} to derive $\eta/s$. 
To our surprise, we find that when $b=30$, the strongly coupled scalar system 
can reproduce all thermodynamic and transport properties of hot quark-gluon system
near $T_c$. $p_T/\epsilon_T$ at ${T_c}$ is close to the lattice QCD result $0.07$,  
$\Delta/d=0.48$ ($d=1$ for scalar system) at $T_c$ is close to the lattice result of the 
peak value $\Delta_{LAT}^{Nf=2}/(d_G+d_Q)\simeq0.4$ at $T_{max}$. The bulk viscosity
to entropy density ratio $\zeta/s$ at $T_c$ is around $0.5\sim 2.0$, which agrees well with
the lattice result in Ref. \cite{LAT-xis-Meyer}. (Note, here $\zeta/s=0.5, 2$ correspond to 
$\omega_0=10 T, 2.5 T$, respectively.) More surprisingly, the shear viscosity over entropy 
density ratio $\eta/s$ at $T_c$ is $0.146$, which also beautifully agrees with lattice 
result $0.1\sim 0.2$ in Ref. \cite{LAT-etas}. In Table \ref{table-all}, we compare our 
results of equation of state and transport 
properties in scalar field theory at $T_c$, and corresponding results in
lattice QCD calculations \cite{LAT-xis-KT,LAT-xis-Meyer,LAT-EOS}, 
the Polyakov-loop Nambu--Jona-Lasinio (PNJL) model \cite{EOS-PNJL-Weise,EOS-PNJL-Ray}, 
and black hole duals \cite{Gubser-EOS}. 

\begin{table}[th]\vspace*{-0.0cm}
\begin{center}%
\begin{tabular}
[c]{|c|c|c|c|c|c|}\hline
  & $\eta/s$ & $\zeta/s$ & $\Delta/d$ & $ c_s^2$ & $p_T/\epsilon_T$ \\\hline
$b=0.3$ & $2065$ & $\simeq 0$ & $0.018$ &  $0.312$  & $0.32$ \\\hline
$b=30$  & $0.146 $ & $0.5\sim 2.0$ & $0.48$ & $0.03$ & $0.07$ \\\hline
${\rm LAT}_{G}$ & $0.1\sim 0.2$ & $0.5\sim 2.0$ & $0.25$ & $-$ & $0.07$ \\ \hline
${\rm LAT}_{Nf=2}$ &  $-$ & $0.25\sim 1.0$ & $0.4$ & $0.05$ & $0.07$ \\ \hline
${\rm PNJL}$ &  $ - $ & $-$ & $0.21$ & $0.08$ & $0.075$ \\ \hline
${\rm AdS/CFT}$ &  $ 1/4\pi $ & $0$ & $0$ & $1/3$ & $1/3$ \\ \hline
${\rm Type I BH}$ &  $ 1/4\pi $ & $0.06$ & $-$ & $0.05$ & $-$ \\ \hline
${\rm Type II BH}$ &  $ 1/4\pi $ & $0.08$ & $-$ & $\simeq 0$ & $-$ \\ \hline
\end{tabular}
\end{center}
\vspace*{-0.1cm}
\caption{Thermodynamic and transport properties at $T/T_c=1$ in scalar theory at weak 
coupling $b=0.3$ and strong coupling $b=30$, in lattice QCD 
\cite{LAT-xis-KT,LAT-xis-Meyer,LAT-EOS}, PNJL model 
\cite{EOS-PNJL-Weise,EOS-PNJL-Ray}, and black hole dules \cite{Gubser-EOS}. The
degeneracy factor $d=1$ for real scalar model. }%
\label{table-all}%
\vspace*{-0.15cm}
\end{table}

In summary, in the Hartree approximation of CJT formalism, we have investigated the equation 
of state and transport properties of the real scalar field model with $Z(2)$ symmetry breaking 
in the vacuum and 2nd-order phase transition at finite temperature. 
We have seen that at phase transition, the system either in weak coupling or strong coupling shows
some common properties: 1)  $p_T/\epsilon_T$, the square of the speed of sound $c_s^2$ as well as
$\eta/s$ exhibit downward cusp behavior at $T_c$. 2) The trace anomaly $\Delta$, the specific heat
$C_v$ as well as $\zeta/s$ show upward cusp behavior at $T_c$. The cusp behavior is related to the 
biggest change rate of entropy density at $T_c$.
At weak coupling, the scalar system near phase transition is asymptotically conformal. However,  
the shear viscosity is huge. At strong coupling, the scalar system near phase transition is highly
non-conformal, the shear viscosity is small, but the bulk viscosity is large. We can expect that for any 
nonconfomal field theory, the corresponding system at strong coupling
near phase transition exhibits highly nonconformality. AdS/CFT method maybe cannot help us 
understand the strongly interacting quark gluon plasma. Unexpectedly, the simplest scalar field model can
do the job. We have found that lattice QCD results on the equation of state and transport
properties near phase transition can be amazingly very well described by the simplest real
scalar model at strong coupling when $b=30$. It is urgent to include the bulk viscosity correction and the 
nonconformal equation of state in hydrodynamics to investigate hadronization and 
freeze-out processes of QGP created at heavy ion collisions 
\cite{bulk-Mishustin,bulk-Muller}, it would be also interesting to investigate
how bulk viscosity affects charm radial flow \cite{Xuzhangbu} at RHIC.

{\bf Acknowledgments.} This work is supported by CAS program
"Outstanding young scientists abroad brought-in", CAS key project KJCX3-SYW-N2, 
NSFC10735040, NSFC10875134 and NSFC10675077.


\end{document}